\documentclass[12pt,a4paper]{article}
\usepackage{amsmath,amssymb,latexsym} 

\newcommand{\muk}{\mu^+ K^0}
\newcommand{\epi}{e^+\pi^0}
\newcommand{\et}{e^+\eta}
\newcommand{\ek}{e^+K^0}
\newcommand{\mpi}{\mu^+\pi^0}
\newcommand{\mt}{\mu^+\eta}
\newcommand{\ero}{e^+\rho^0}
\newcommand{\eo}{e^+\omega}
\newcommand{\eks}{e^+K^{*0}}
\newcommand{\mro}{\mu^+\rho^0}
\newcommand{\mo}{\mu^+\omega}
\newcommand{\nep}{\bar\nu_e\pi^+}
\newcommand{\nek}{\bar\nu_e K^+}
\newcommand{\nmpi}{\bar\nu_\mu\pi^+}
\newcommand{\nmk}{\bar\nu_\mu K^+}
\newcommand{\nero}{\bar\nu_e\rho^+}
\newcommand{\neks}{\bar\nu_e K^{*+}}
\newcommand{\nmro}{\bar\nu_\mu\rho^+}
\newcommand{\nmks}{\bar\nu_\mu K^{*+}}
\newcommand{\ntp}{\bar\nu_\tau\pi^+}
\newcommand{\ntk}{\bar\nu_\tau K^+}
\newcommand{\ntro}{\bar\nu_\tau\rho^+}
\newcommand{\ntks}{\bar\nu_\tau K^{*+}}
\newcommand{\epin}{e^+\pi^-}
\newcommand{\mpin}{\mu^+\pi^-}
\newcommand{\eron}{e^+\rho^-}
\newcommand{\mron}{\mu^+\rho^-}
\newcommand{\nepn}{\bar\nu_e\pi^0}
\newcommand{\nekn}{\bar\nu_e K^0}
\newcommand{\nmpin}{\bar\nu_\mu\pi^0}
\newcommand{\nmkn}{{\bar\nu}_\mu K^0}
\newcommand{\neron}{\bar\nu_e\rho^0}
\newcommand{\neksn}{\bar\nu_e K^{*0}}
\newcommand{\nmron}{\bar\nu_\mu\rho^0}
\newcommand{\nmksn}{\bar\nu_\mu K^{*0}}
\newcommand{\neon}{\bar\nu_e \omega}
\newcommand{\nmon}{\bar\nu_\mu \omega}
\newcommand{\netn}{\bar\nu_e \eta}
\newcommand{\nmtn}{\bar\nu_\mu \eta}
\newcommand{\ntpn}{\bar\nu_\tau \pi^0}

\newcommand{\ntkn}{\bar\nu_\tau K^0}
\newcommand{\nttn}{\bar\nu_\tau \eta}
\newcommand{\ntron}{\bar\nu_\tau \rho^0}
\newcommand{\nton}{\bar\nu_\tau \omega}
\newcommand{\ntksn}{\bar\nu_\tau K^{*0}}

\newcommand{\bye}{\end{document}}

\newcommand{\be}{\begin{equation}}
\newcommand{\ee}{\end{equation}}
\newcommand{\beq}{\begin{eqnarray}}
\newcommand{\eeq}{\end{eqnarray}}
\newcommand{\plr}{$P_{LR}$}

\newcommand{\lra}{\longrightarrow}

\begin{document}

\def\NPB#1#2#3{Nucl. Phys. {\bf B} {\bf#1} (19#2) #3}
\def\PLB#1#2#3{Phys. Lett. {\bf B} {\bf#1} (19#2) #3}
\def\PRD#1#2#3{Phys. Rev. {\bf D} {\bf#1} (19#2) #3}
\def\PRL#1#2#3{Phys. Rev. Lett. {\bf#1} (19#2) #3}
\def\PRT#1#2#3{Phys. Rep. {\bf#1} C (19#2) #3}
\def\ARAA#1#2#3{Ann. Rev. Astron. Astrophys. {\bf#1} (19#2) #3}
\def\ARNP#1#2#3{Ann. Rev. Nucl. Part. Sci. {\bf#1} (19#2) #3}
\def\MODA#1#2#3{Mod. Phys. Lett. {\bf A} {\bf#1} (19#2) #3}
\def\NC#1#2#3{Nuovo Cim. {\bf#1} (19#2) #3}
\def\ANPH#1#2#3{Ann. Phys. {\bf#1} (19#2) #3}
\def\PROG#1#2#3{Prog. Th. Phys. {\bf#1} (19#2) #3}  
\begin{titlepage}
\setcounter{page}{1}

\begin{large}
\title{Are Right-Handed Mixings Observable?~\footnote{Talk at the
  ``Corfu Summer Inst. on Elementary Particle Physics'', Sept. 1998.}} 
\author{Yoav Achiman \\[0.5cm]
        Department of Physics\\
        University of Wuppertal\\
        Gau\ss{}str.~20, D--42097 Wuppertal \\
        Germany\\[1.5cm]}

\date{December 1998}

\maketitle
\setlength{\unitlength}{1cm}
\begin{picture}(5,1)(-12.5,-12)
\put(0,0){WUB 98-43}
\put(0,1){hep-ph/9812389}
          
\end{picture}

\begin{abstract} 
Asymmetric mass matrices can induce  large RH mixings. Those are non
-measurable in the  SM but are there and play an important role in its
extensions. The RH rotations are in particular relevant  for the proton decay, 
neutrino properties and baryon asymmetry. E.g. large RH mixings lead 
to kaon dominated proton decay even without SUSY  and could be the reason 
for a large neutrino mixing. By studying those phenomena one can learn about
the RH rotation matrices and this can 
reduce considerably the arbitrariness  in the present fermionic mass study.
\end{abstract} 

\thispagestyle{empty}
\end{large}

\end{titlepage}
\clearpage

Right-handed  (RH) mixings are not  relevant in the framework  of the
standard model (SM). Also, RH currents have not been observed 
experimentally (yet?).
So, why are RH mixings interesting?
 
What are RH mixings?\\
To  diagonalize   a general complex (mass) matrix M one needs
a bi-unitary transformation, i.e.
two unitary  matrices $U_{L,R}$, such that
\begin{equation}
{U_L}^\dagger  M U_R = M_{diagonal}
\end{equation}
or
\begin{equation}
{U_L}^\dagger MM^\dagger U_L = (M_{diag.})^2 = {U_R}^\dagger M^\dagger MU_R .
\end{equation}
 
Only in the case of hermitian (symmetric) matrices is $U_R$ related 
to $U_L$
\begin{equation}
M=M^\dagger (M^T) \ \Longrightarrow \  U_R=U_L({U_L}^*) .
\end{equation}
 
RH  fermions are singlets in the SM and only LH charged currents
are involved in the weak interactions
\begin{equation}
{\cal L}_W = {W^\dagger}_\mu{\overline {u_L}} \gamma ^\mu V_{CKM} d_L
\  
+ \   h.c.
\end{equation}
where
$$
V_{CKM} = {U_L^u}^\dagger U_L^ d  .
$$
 
The $U_R$'s do not play a role in the SM. However, the fermionic mass 
matrices are generated here by unknown Yukawa couplings and 
therefore are completely arbitrary. Hence, the SM must be extended  to 
``explain''
the fermionic masses and mixings, an extension which is already  
suggested by
\begin{itemize}
\item{Grand Unification:      
$\alpha_1(M_W) , \alpha_2(M_W) , \alpha_3(M_W) \
\rightarrow \ \alpha(M_{GUT})$}
\item{Yukawa Unification:  $m_\tau(M_{GUT})  \simeq  m_b(M_{GUT}) $}
\item{ L-R restoration at   $M_{R}  \gg M_W$}
\item{Mixed massive neutrinos (seesaw)~\cite{see} with: $M_{\nu_R}  \gg
M_W$     etc.}.
\end{itemize}
Many different ``models'' are known to give the right masses of the 
charged fermions and $V_{CKM}$ (within the experimental 
errors)~\cite{rrr}~\cite{as} and this is an indication that the mass
problem is far from being solved. Part of this freedom is due to the 
fact that these suggestions disregard the RH rotations.

Most models use hermitian mass matrices for no other reasons 
than simplicity\cite{rrr}. However, recently more and more asymmetric mass
matrices are used (mainly to have additional freedom for the neutrino 
sector)\cite{as}. Asymmetric mass matrices imply $U_L \not=  U_R$,  so that
here the $U_R$'s are a clue to distinguish between different models. 

It is true that  RH currents  have not been observed till
now\footnote{There is a certain indication that RH currents can be
observed in bottom decays.~\cite{RH}}  but this means only that the relevant
gauge bosons are heavy and/or mix very little with the observed LH ones
and/or the RH neutrinos are very heavy.
The limits on RH gauge bosons are clearly very model
dependent~\cite{RH}.

Our main point is however that even if RH currents will not be directly
observed at low energies they play an important  role at energies where 
the L-R  symmetries are restored. RH mixings effect therefore phenomena 
like:
\begin{itemize}
\item{Proton decay}
\item{Neutrino seesaw~\cite{see}}
\item{Leptogenesis via decays of RH neutrinos as the origin of baryon 
asymmetry~\cite{bar}  etc. },
\end{itemize}
which are indirectly observable.

Now, it is clear that 
the symmetries which dictate  the mass matrices are effective at scales
relevant for the theories beyond the SM. In those theories the
RH mixings are not arbitrary any more, there are  also no reason to
assume that they are small. 
Actually even large RH mixings are not unnatural and are the standard
in $P_{LR}$ invariant theories~\cite{plr} 
We claim also that the large leptonic
mixing (recently observed  by Super-Kamiokande~\cite{sk}) may be 
related to large RH rotations.

What is $P_{LR}$ ?\\
In the framework of Current Algebra it is common to assign the
baryons to a $P$-~invariant \ $(3,{\bar 3}) \oplus ({\bar
3},3)$ \  representation under the {\em global} chiral group:\\
\centerline{$SU_L(3) \times SU_R(3) \times P$~\cite{GM}.}\\
The baryons  acquire their masses when the chiral group is broken  into its
diagonal subgroup
$SU_{L+R}(3)$\ , under which the baryons constitute \  ${\bf 8} \oplus
{\bf 1}$ \  Dirac spinors.

An analogous symmetry can be applied to fermions in  $L-R$
symmetric gauge theories.
As an example, let us consider the leptons in the $E_6$ GUT~\cite{e6}.
Those are LH  Weyl spinors that transform like  $(1,3,{\bar 3})$
under the  maximal subgroup of $E_6$,
$$
E_6 \supset SU_C(3) \times SU_L(3) \times SU_R(3)\quad .
$$
 Whereas  $P$-reflection for the global symmetry leads per definition to
$SU_L(3) \leftrightarrow \nobreak SU_R(3)$ exchange, in the gauge theories
$L,R$ are only an historical notation. The chirality of the local currents
is fixed by the representation content of the fermions under
 $SU_L(3) \times SU_R(3)$\ . Hence, for gauge theories we have to
require, in addition to Parity exchange, also $SU_L(3) \leftrightarrow \nobreak SU_R(3)$.
The irreducible representation  of the leptons under  
$SU_C(3) \times SU_L(3) \times SU_R(3) \times P_{LR}$ \ is 
$$
(1,3,{\bar 3})_{LH} \oplus  (1,{\bar 3},3)_{RH} \quad ,
$$
which requires two families.
 
Under the diagonal \  $SU_C(3) \times SU_{L+R}(3)$ \  one obtains then
 \ ${\bf 8} \oplus {\bf 1}$ \  of Dirac spinors. Applying this to the $e$ and
$\mu$ families this is realized in analogy
with the hadrons as follows.

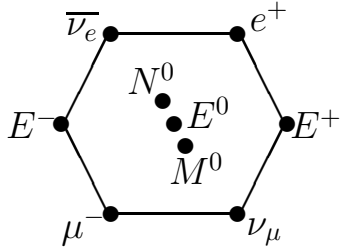
\begin{center}
\thicklines
\setlength\unitlength{0.3mm}
\begin{picture}(70,70)(-75,-35)
\put(0,0){\circle*{7}}
\put(-5,10){\circle*{7}}
\put(5,-10){\circle*{7}}
\put(28,40){\circle*{7}}
\put(-28,40){\circle*{7}}
\put(-28,-40){\circle*{7}}
\put(28,-40){\circle*{7}}
\put(-50,0){\circle*{7}}
\put(50,0){\circle*{7}}
\put(-25,40){\line(1,0){50}}
\put(-25,-40){\line(1,0){50}}
\put(-48,2){\line(1,2){18}}
\put(30,-38){\line(1,2){18}}
\put(-30,-38){\line(-1,2){18}}
\put(48,2){\line(-1,2){18}}
\put(45,54){\makebox(0,0)[t]{\large $e^+$ }}
\put(-38,50){\makebox(0,0)[t]{\large $\overline {\nu_e}$ }}
\put(-40,-54){\makebox(0,0)[b]{\large $\mu^-$}}
\put(40,-54){\makebox(0,0)[b]{\large $\nu_\mu$}}
\put(63,0){\makebox(0,0){\large\bf $E^+$}}
\put(-63,0){\makebox(0,0){\large\bf $E^-$}}
\put(15,5){\makebox(0,0){\large\bf $E^0$}}
\put(-10,20){\makebox(0,0){\large\bf $N^0$}}
\put(10,-20){\makebox(0,0){\large\bf $M^{0}$}}
\end{picture}
\end{center}  

Such a model was actually constructed in 1977\cite{Pl}  when the third heavy
family was not yet observed. It is quite a general belief now that this
top-family is the only one acquiring masses through direct coupling to the
Higgs representation, while the light families get their masses through second order
``corrections''. It is then natural that these two light families obey
symmetries like \ $P_{LR}$ \ .  When those symmetries are broken, the
particles gain their physical masses and mixings.\footnote{We know that
in SUSY theories as well, sfermions of the two light families must be
quite degenerate to avoid FCNCs.}
 
The \  \plr \  operation can be formally defined in terms of two families
\cite{plr}
\begin{equation}
P_{LR} \ f^i(x) \  P^{-1}_{LR} = \epsilon^{ij} \sigma_2 \hat{f}^{j\star}
(\bar x) \quad .
\end{equation}
The \  \plr \  invariant Lagrange looks then as follows 
\begin{equation}
{\cal L}_Y = y_{12}  \overline{\Psi^{1c}} \Phi_{12} \Psi^2  -
                      y_{21}  \overline{\Psi^{2c}} \Phi_{21} \Psi^1  +  h.c.
\end{equation}
The corresponding mass matrices are hence pure off-diagonal in this limit  
$$
\begin{array}{cc}
\begin{array}{cc}
M_2^u =
\left (
\begin{array}{cc}
0&-m_u\\
m_c&0
\end{array} \right ) & 
M_2^d =
\left (
\begin{array}{cc}
0&-m_d\\
m_s&0
\end{array} \right ) 
\end{array} \\ [0.5cm]
\begin{array}{cc} 
M_2^e =
\left (
\begin{array}{cc}
0&-m_e\\
m_\mu&0
\end{array} \right ) & 
M_2^\nu =
\left (
\begin{array}{cc}
0&-m_{\nu_e}\\
m_{\nu_\mu}&0
\end{array} \right ) 
\end{array}.
\end{array}
$$

These matrices can be diagonalized by the transformations
$$
\begin{array}{ccccc}
\left (
\begin{array}{cc}
1&0\\
0&1
\end{array} \right ) &
\left (
\begin{array}{cc}
0&-m_1\\
m_2&0
\end{array} \right) &
\left (
\begin{array}{cc}
0&1\\
-1&0
\end{array} \right) &
=
\left (
\begin{array}{cc}
m_1&0\\
0&m_2
\end{array} \right) 
\end{array}.
$$
and those are equivalent to the exchanges
\begin{equation}
u^c_{LH} \longleftrightarrow c^c_{LH}  \qquad \qquad  d^c_{LH}
\longleftrightarrow s^c_{LH}  \qquad \qquad  e^+_{LH} \longleftrightarrow
\mu^+_{LH} ,
\end{equation}
 
which mean {\em full} RH {\em rotations}. Applying this to the effective dim.6
B-violating Lagrangian of $SO(10)$\cite{lang} 
and noting that only the two light 
families are relevant for the proton decay, two decay modes result~\cite{key}
$$
P \lra  \nmk  \qquad \hbox{and}  \qquad  P \lra \muk .
$$
Now, to make such a model realistic one must break \plr \  by a small
amount, to allow for Cabbibo mixing and add the heavy t-family. Also, 
to induce gauge unification (without SUSY) an intermediate breaking
scale,  $M_I \approx  10^{12}$ GeV is required. This is however also
the right RH neutrino mass scale for the seesaw mechanism~\cite{see}  
and leptogenesis~\cite{bar} as well as the
scale of the invisible Axion window~\cite{ax}.
  
In this talk I would like to report on a systematic study of models
with large RH rotations and their possible effects.
I will give an example in terms of  a ``realistic'' $SO(10)$ Model with such
mixings. By this I mean a conventional $SO(10)$ theory that reproduces
all the observed fermionic 
masses and LH mixings but at the same time generates large RH angles.\\
This can be obtained by requiring small deviations from  the \plr \
invariant case. E.g.  consider  at the high unification scale the 
following mass matrices
(those can be obtained using  a global $ U_f(1)$ or a discrete  
symmetry)\cite{am}

$$
\begin{array}{ccc}
m_d=
\left (
\begin{array}{ccc}
0&-m_d&0\\
m_s&0&0\\
0&0&m_b
\end{array} \right ) & 
m_u=
\left (
\begin{array}{ccc}
a&m_1&b\\
m_2&0&0\\
c&0&m_3
\end{array} \right ) &  
m_\ell=
\left (
\begin{array}{ccc}
0&-m_e&0\\
m_\mu&0&0\\
0&0&m_c
\end{array} \right )  
\end{array}.
$$ 

These matrices give the following RH angles,in the u-sector, at the high scale
 
\begin{equation} 
\begin{array}{ccc}
\Theta_{12}^R = 1.57 \  rad. &  \Theta_{23}^R = 0.0 \  rad. & \Theta_{13}^R = 
-1.50 \  rad. 
\end{array} 
\end{equation}
 
We  studied in detail the embedding of those matrices in the framework of an 
$SO(10)$ model broken  at $M_U$ to the Pati-Salam group~\cite{LRD} 
and this in the 
second step to the SM at $M_I$ 
\begin{equation}
SO(10){}\stackrel{M_U}{\longrightarrow} {}SU_C(4) \times SU_L(2) \times SU_R(2)
\  \stackrel{M_I}{\longrightarrow} \  SM
\end{equation}
 
The  Higgs representations needed for the local breaking  and the
generation of the fermionic mass 
matrices, fix the two loop  renormalization group equations (RGEs). Those are 
used for two  cases,
one with D-Parity ($g_L=g_R$) and the other without it  ($g_L \neq
g_R$). We found:\\
 
with D-Parity:
\begin{equation}
\begin{array}{ccc}
M_U = 1.04 \times 10^{15} GeV & M_I = 5.66 \times 10^{13} GeV  & \alpha_U = 
0.02841
\end{array}
\end{equation}

and without D-Parity:
 
\begin{equation}
\begin{array}{ccc}
M_U = 5.68 \times 10^{15} GeV & M_I = 2.09 \times 10^{11} GeV
& \alpha_U = 0.04207 
\end{array}
\end{equation}
 
Using  then the fermionic mass matrices  and  $V_{CKM}$ at $M_Z$  we  
evaluated the values of the matrix elements at $M_I$ and also give the
RH mixing angles at this scale.   Those values were used to 
calculate the proton and neutron B-violating branching ratios (see
tab. 1 and tab. 2).\\

\begin{table}[h]
\begin{center}
\begin{tabular}{|l|r||l|r|}
\hline
\quad channel & ratio (\%) & \quad channel & ratio (\%) \\
\hline
\hline
\quad $\epi$ & 0.0 \; & \quad $\nep$  & 0.0 \; \\
\quad $\ek$  & 3.6 \; & \quad $\nek$  & 0.0 \; \\
\quad $\et$  & 0.0 \; & \quad $\nmpi$ & 0.0 \; \\
\quad $\mpi$ & 2.6 \; & \quad $\nmk$  &56.2 \; \\
\quad $\muk$ &27.6 \; & \quad $\nero$ & 0.0 \; \\
\quad $\mt$  & 0.5 \; & \quad $\neks$ & 0.0 \; \\
\quad $\ero$ & 0.0 \; & \quad $\nmro$ & 0.0 \; \\
\quad $\eo$  & 0.0 \; & \quad $\nmks$ & 8.0 \; \\
\quad $\eks$ & 0.0 \; & \quad $\ntp$  & 0.0 \; \\
\quad $\mro$ & 0.2 \; & \quad $\ntk$  & 0.0 \; \\
\quad $\mo$  & 1.2 \; & \quad $\ntro$ & 0.0 \; \\
             &        & \quad $\ntks$ & 0.0 \; \\
\hline
\end{tabular}
\begin{minipage}{10.6cm}
\caption{\label{tab3} Branching ratios $\Gamma_i/\Gamma$ for proton
  decay channels (without neutrino mixing);
  total decay rate: 
$\Gamma =~9.4 \ 10_{}^{-35}
  \textrm{yr}_{}^{-1}
  = (1.1 \ 10_{}^{34} \; \textrm{yr})_{}^{-1}$}
\end{minipage}
\end{center}
\end{table}
\begin{table}[h]
\begin{center}
\begin{tabular}{|l|r||l|r|}
\hline
\quad channel & ratio (\%) & \quad channel & ratio (\%) \\
\hline
\hline
\quad $\epin$  & 0.0 \; & \quad $\neon$  & 0.0 \; \\
\quad $\mpin$  & 3.8 \; & \quad $\neksn$ & 0.0 \; \\
\quad $\eron$  & 0.0 \; & \quad $\nmron$ & 0.0 \; \\
\quad $\mron$  & 0.2 \; & \quad $\nmon$  & 0.0 \; \\
\quad $\nepn$  & 0.0 \; & \quad $\nmksn$ & 3.9 \; \\
\quad $\nekn$  & 0.0 \; & \quad $\ntpn$  & 0.0 \; \\
\quad $\netn$  & 0.0 \; & \quad $\ntkn$  & 0.0 \; \\
\quad $\nmpin$ & 0.0 \; & \quad $\nttn$  & 0.0 \; \\
\quad $\nmkn$  &92.1 \; & \quad $\ntron$ & 0.0 \; \\
\quad $\nmtn$  & 0.0 \; & \quad $\nton$  & 0.0 \; \\
\quad $\neron$ & 0.0 \; & \quad $\ntksn$ & 0.0 \; \\
\hline
\end{tabular}
\begin{minipage}{10.6cm}
\caption{\label{tab4} Branching ratios $\Gamma_i/\Gamma$ for neutron
  decay channels (without neutrino mixing);
  total decay rate: 
$\Gamma =~1.3 \ 10_{}^{-34}
  \textrm{yr}_{}^{-1}
  = (7.8 \ 10_{}^{33} \; \textrm{yr})_{}^{-1}$}
\end{minipage}
\end{center}
\end{table}
%

We obtained very similar results in those two cases and only the
absolute rates depend on the details of the local breaking.\\
Without D-Parity we obtain:
\begin{equation}
\tau_{total}^{proton} = 1.1\times 10^{34\pm.7\pm1.0^{+.5}_{-5.0}}\quad yrs.  
\end{equation}
 
For the uncertainties and threshold corrections we used the estimates
of Langacker~\cite{lang} and Lee~{\em et al}~\cite{par}. 
 
Our main prediction are the branching ratios which are independent on those 
uncertainties and the details of the local breaking. The absolute
rates indicate, however, that the results of the model are well in the 
range of observability of the new proton decay experiments~\cite{Pex}.
The branching ratios are very similar to the ``smoking gun''
predictions of the 
SUSY GUTs~\cite{sgut} and in contradiction with the conventional GUTs where  
$P \lra  \epi  \quad$ dominates.
Using a $U(1)_F$  one can obtain naturally large leptonic mixings induced by 
the large RH  rotations~\cite{am}.
We will study also effects of large RH mixings on  the proton decay in SUSY 
SO(10). Those could play an important role in view of the fact that it
was shown 
recently that RRRR and RRLL effective dim.5  operators can dominate
proton decay in such models~\cite{RR}. 
Also, effects of SUSY and non SUSY leptogenesis as the origin of the baryon 
asymmetry~\cite{bar} will be considered.

Part of this work was done in collaboration with Carsten  Merten. 
I would like to thank also M.~K.~Parida for discussions and for
pointing to us a mistake.

\end{document}